\documentclass[%
 reprint, nofootinbib,
 amsmath,amssymb,
 aps,
]{revtex4-2}

\usepackage[dvipsnames]{xcolor}
\usepackage{appendix}
\usepackage[utf8]{inputenc}
\usepackage{soul}
\usepackage{graphicx}
\usepackage{mathrsfs}
\usepackage{hyperref}
\usepackage{braket}
\usepackage{amsthm}
\usepackage{float}
\usepackage{notoccite} 
\hypersetup{
    colorlinks=true,
    linkcolor=magenta,
    citecolor = blue,
    filecolor=magenta,      
    urlcolor=blue,
    pdftitle={Overleaf Example}
    }
\usepackage{media9}
\usepackage{float}
\usepackage{dirtytalk}
\usepackage{mathrsfs}
\usepackage{amsmath,amssymb}
\usepackage{physics}
\usepackage{tikz}
\usetikzlibrary{quantikz}
\usepackage[export]{adjustbox}
\usepackage{lipsum} 
\graphicspath{{Figures/}}
\usepackage{caption}
\usepackage{mdframed}
\usepackage[font=normal,labelfont=bf]{caption}
\usepackage[caption=false]{subfig}
\usepackage{mathtools}
\begin{document}

\preprint{APS/123-QED}

\title{Decoherence due to Spacetime Curvature}

\author{Raghvendra Singh}
\email{raghvendra@imsc.res.in}
\affiliation{%
Institute of Mathematical Sciences, Homi Bhabha National Institute,
C.I.T. Campus, Chennai 600 113.
}%


\author{Kabir Khanna}
\email{kabir.khanna@wadham.ox.ac.uk} 
\affiliation{
Mathematical Institute, University of Oxford, Oxford, United Kingdom.
\\
\&
\\
Department of Engineering Design, Indian Institute of Technology Madras, Chennai 600 036.
}%

\author{Dawood Kothawala}
\email{dawood@iitm.ac.in}
\affiliation{Centre for Strings, Gravitation and Cosmology, Department of Physics, Indian Institute of Technology Madras, Chennai 600 036.}


\date{\today}

\begin{abstract}
\noindent There has been considerable interest over the past years in investigating the role of gravity in quantum phenomenon such as entanglement and decoherence. In particular, gravitational time dilation is believed to decohere superpositions of center of mass of composite quantum systems. 
Since true effects of gravity are encoded in the curvature of spacetime, the universality of such decoherence must be characterized through components of Riemann tensor $R_{abcd}$, with a clear separation from non-inertial kinematic effects. We obtain the reduced density matrix of a composite system in a generic curved spacetime and express the decoherence time scale explicitly in terms of curvature. The decoherence in an inertial frame is caused by tidal acceleration. We also analyze the effects of self-gravity and show that the coupling of gravitational interaction with external curvature can not be captured by the replacement $m \to m + H_{\rm int}/c^2$.
\end{abstract}


\maketitle


\section{Introduction}
The interplay between gravitation and quantum mechanics in extreme physical situations involving high energies and/or strong gravitational fields is typically believed to yield new physics not accessible to existing observations and experiments. However, it has become evident in the past decade or so that the interplay between gravity and low-energy quantum systems is interesting in its own right \cite{Pikovski:2013qwa, Zych:2012ut, Zych:2011hu, Zych:2015wfk, Zych:2017tau, Zych:2015fka, Pikovski:2015wwa}. This regime can be well approximated by the framework of relativistic quantum mechanics in its first quantized form on a background spacetime. When gravity is itself treated classically, the background spacetime can be taken as some exact solution of Einstein equations which is simple enough for the analysis to be tractable and yield analytical results. This is often not possible, and even when it is, does not yield useful insights since the choice of a specific solution hides the manner in which Riemann curvature explicitly appears in and affects the final results. An alternative route is to employ a suitable frame in which kinematic and curvature effects can be captured cleanly in terms of physically observable quantities. This latter approach provides remarkable insights when used to study quantum and thermodynamic properties of systems in curved spacetimes, as well as their connection with Einstein equations \cite{Parker:1980hlc, Kothawala:2011fm, Kothawala:2016lge}. In fact, recent work has shown that the latter approach also has the potential to yield new insights into some non-perturbative effects of spacetime curvature on behaviour of classical and quantum probes \cite{K:2021gns}, as well as implications for curvature for the uncertainty principle at high energies \cite{Singh:2021iqa}. 

In this work, our focus will be on how spacetime curvature can affect the quantum mechanical properties of a composite system. One would generically expect the two key features of quantum mechanics and general relativity -- time dilation, and quantum superposition and entanglement -- to play an inevitable role in such an analysis. A well-known result in this context is by Pikovski et al. \cite{Pikovski:2013qwa} (PZCB), which shows that gravitational time dilation will generically cause decoherence of the center-of-mass superpositions for a composite system. This happens essentially because the background metric indirectly couples the internal degrees of freedom with the center of mass, through the total Hamiltonian, thereby producing a reduced density matrix for the latter that exhibits decoherence. Our aim
will be to describe this quantum decoherence directly in terms of curvature of the background spacetime, so as to provide a clear separation between effects arising due to the non-inertial nature of the frame from those produced due to curvature. It is the latter that captures the true effects due to gravity. To achieve this, we will introduce a fully covariant setup for discussing this problem, thereby removing any ambiguities related to the choice of coordinates, etc. 

Three main results that we establish in this work are: 
\begin{enumerate}
    \item The decoherence time scale derived in PZCB can be obtained purely in terms of the electric part of the Riemann tensor in a given frame.
    \item The magnetic part of Riemann alters the PZCB by introducing an additional term. 
    \item The gravitational self-interaction itself does not couple to the external curvature through the simple replacement $m \to m + H_{\rm int}/c^2$ in the point particle Hamiltonian.
\end{enumerate}

%
 In the first part, we consider a setup similar to one discussed by PZCB in \cite{Pikovski:2013qwa}, but in an arbitrary curved background. We take the system to be a composite system with some internal degrees of freedom in an arbitrary background, whose center of mass degree of freedom is in a superposition of two suitably defined position eigenstates $\ket{x_1}$ and $\ket{x_2}$ at some initial time ($t_1=0$ surface in the figure (\ref{fig:FNC})). The internal dynamics is governed by the Hamiltonian $H_{\rm int}$, whose coupling to curvature is fixed by assuming it contributes to the inertial mass as $H_{\rm int}/c^2$. We then follow the conventional route of computing the density matrix at a later time, tracing over the internal degrees of freedom and finding the reduced density matrix for the center of mass. The interferometric Visibility so obtained is reduced purely due to gravitational effects, even when special relativistic effects are absent. This then gives us results (1) and (2) above.

In the second part, we consider the case where the system itself distorts the background geometry, thereby incorporating the effect of self-gravitation. To do this, we consider the highly simplified case of a system of two masses, $M$ and $m$, with $M\gg m$, placed in an external curved spacetime. Assuming $m$ to be moving in the effective geometry corresponding to background curvature + curvature due to $M$, we determine the coupling of the \textit{gravitational interaction} -- $(- GMm/r)$ -- with the background curvature. This coupling turns out to be different from the manner in which other, non-gravitational, interactions are coupled to the inertial mass through an application of the equivalence principle. This will be our result (3), and we discuss its implications for the strong equivalence principle in detail in Section (\ref{Self-gravity}).     

\section{Quantum systems in curved spacetime}
\label{sec2}
\begin{figure}[H]%
    {{\includegraphics[height=9cm, width=9.3cm]{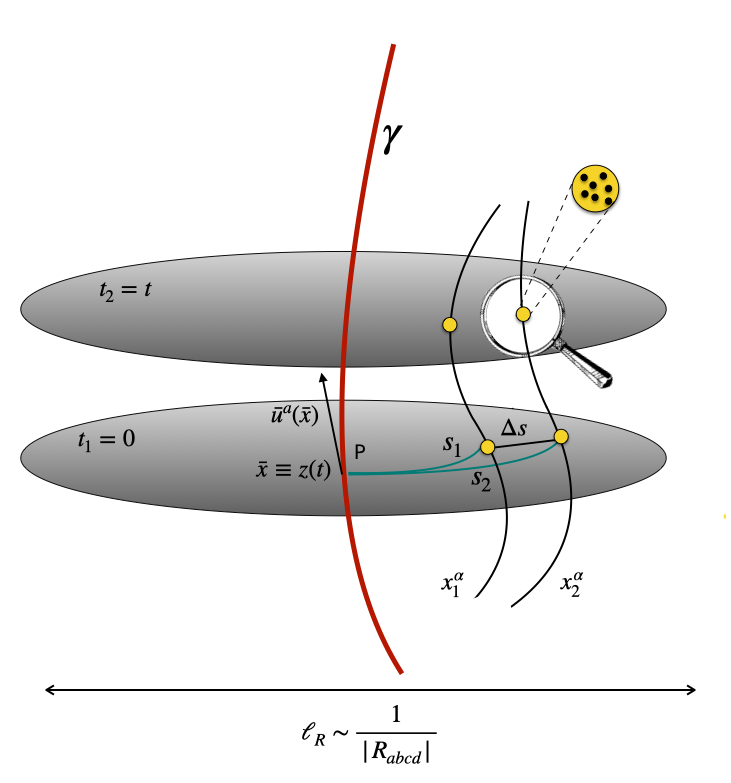} }}%
    \caption{Description of a composite system (yellow spheres) in a reference frame characterized by a timelike curve $\gamma$ (the thick red curve). The center of mass is supposed to be in a quantum superposition of two trajectories (black curves). A relational description of this system in arbitrary curved spacetime is provided by Fermi normal coordinates, in which the coordinates, as well as the separation between the superposed states, measure the deviation between the curves in a covariant manner. (See text for details.)  
    %
    }%
    \label{fig:FNC}%
\end{figure}

To understand the curvature effects on the quantum interference, we consider the arbitrary curved background with curvature $R_{abcd}$ and set up the calculation in a frame characterized by a timelike curve $\gamma$. Consider a system whose center of mass is in the superposition of the two distinct spatial positions $x_1^{\alpha}$ and $x_2^{\alpha}$ where $ x ^{\alpha}=-e^{\alpha}_{\bar a}(\bar x) \sigma^{\bar a}(x, \bar x)$ with $(\bar u^a(\bar x), e^{\alpha}_{\bar a}(\bar x) )$ to be orthonormal tetrad that is propagated along $\gamma$ by Fermi-Walker transport, and $\sigma^{\bar a}(x, \bar x)$ is the gradient of the so-called Synge's world function $\sigma(x, \bar x)$ at the base point $\Bar{x}$. We locate the unique spacelike geodesics $s_1$, and $s_2$ passing through $x_1, x_2$, respectively, and intersect $\gamma$ orthogonally by satisfying the relation $\sigma_{\bar a}(x, \bar x)u^{\bar a}=0$. It should be evident that this entire construction is covariant and, at this stage, independent of any kind of Taylor expansions, etc. However, for explicit computations, we note that the variables employed above are precisely the Fermi normal coordinates \cite{Manasse:1963zz} based on $\gamma$, and we will use the known form of the metric in these coordinates for our ensuing discussion and computation of quantum dynamics as well as decoherence due to curvature. Let $M$ be the mass of the composite system, and the time evolution of its internal degrees of freedom be governed by the internal Hamiltonian $H_{\rm int}$. The internal Hamiltonian does not depend on the position and momentum of COM degree of freedom as it operates in a different Hilbert space. The system dynamics is governed by the Hamiltonian $\hat{H}=\hat{H}_{\rm Free}+\hat{H}_{int}+\hat{H}_{\rm coupling}$. If $\rho_0$ is the initial  for the whole system, the time evolution of the density matrix will be $\rho_t = U \rho_0 U^{\dag}$, where $U$ is the unitary operator.  To see the effect of gravity on quantum interference, we start by choosing  $\gamma$ to be the reference curve for which the center of mass is in the superposition of two semi-classical paths $x_1^{\alpha}$ and $x_2^{\alpha}$ in spacetime with an arbitrary metric as shown in the figure (\ref{fig:FNC}). Since the time evolution of each superposed path depends on the path the system takes by virtue of the metric that generically varies in space, it leads to gaining which-way information for the paths leading to decoherence. The measure of decoherence is mathematically represented as the interferometric Visibility; the value less than unity represents the loss of quantum coherence. The expression for Visibility can be written as
\begin{eqnarray}
    V= \left|Tr\left[e^{- i \int \hat{H} dt}\rho_0 e^{ i \int \hat{H} dt} \right]\right|
\end{eqnarray}
where $\hat{H}$ is the total Hamiltonian, $\rho_0$ represents the initial density matrix for the whole system (internal $\otimes$ COM) at $t=0$, and $Tr$ represents trace over all internal modes. Substituting the expression for the Hamiltonian from the Eq. (\ref{H}), $H_{\rm Free}$ will give a constant phase factor, so it will not contribute to Visibility. If the internal states are not the eigenstates of internal Hamiltonian $H_{\rm int}$, Visibility reduces due to the  terms $H_{\rm int}+H_{\rm coupling}\equiv H_{\rm int}(1+\Theta)$ is
$V= \left|Tr\left[e^{- i \int \hat{H}_{\rm int}(1+\Theta) dt}\rho_i e^{ i \int \hat{H}_{\rm int}(1+\Theta) dt} \right]\right|.
$ 
As illustrated, the off-diagonal terms of the reduced density matrix correspond to the Visibility, which depends on the $\Theta$ function, and $\Theta$ is dependent on the COM variable only 
\begin{widetext}
   \begin{eqnarray} \label{Theta1}
    \Theta=-\frac{\vec p  ^2}{2 m^2 c^2}\left(1+\frac{ a_{\mu} x^{\mu}}{ c^2}+\frac{R_{0 \mu 0 \nu}x^{\mu}x^{\nu}}{2}\right)
    +\frac{1}{2}\left(\frac{2 a_{\mu} x^{\mu}}{ c^2}+R_{0 \mu 0 \nu}x^{\mu}x^{\nu}\right)-\frac{1}{6}R^{\mu \ \nu}_{\ \alpha \ \beta} x^{\alpha}x^{\beta}\frac{p_{\mu}p_{\nu}}{m^2 c^2} 
\end{eqnarray} 
\end{widetext}
For simplification, we consider  slowly moving particles such as $p\ll mc$, $a_{\mu}x^{\mu}\ll c^2$, and ignoring terms of the order $\mathcal{R}x^2\cross (p/mc)^n\; (n> 1)$ since they are of the higher order of smallness, we get
\begin{eqnarray} \label{Theta2}
    \Theta=-\frac{\vec p  ^2}{2 m^2 c^2}+\frac{a_{\mu} x^{\mu}}{c^2}+\frac{R_{0 \mu 0 \nu}x^{\mu}x^{\nu}}{2}.
\end{eqnarray}
The last term in $\Theta$ is the electric part of the tidal tensor and arises due to the curved nature of background geometry. The Visibility will reduce due to the difference between the integrated value of $\Theta$  between two superposition states $\ket{x_1}$ and $\ket{x_2}$ for all moments of time under which the system dynamics runs ($\int_1 ^2 (1+\Theta)dt \equiv \Delta \Theta)$
\begin{eqnarray}\label{Visibility}
     V = |\langle e^{\frac{-i}{\hbar}H_{\rm int} (\Delta \Theta)}\rangle|
\end{eqnarray}

The Visibility expression has the following key features
\begin{enumerate}
    \item The Visibility is unaffected if either $\Delta \Theta= 0$ or if the system is in an eigenstate of the internal Hamiltonian $H_{\rm int}$.
    \item If the system is taken to be subjected by the gravitational potential of the form considered by Pikovski et al. \cite{Pikovski:2013qwa}, only the second term of Eq. (\ref{Theta2}), i.e., $a_{\mu}x^{\mu}\approx g x$, remains. However, this is not a genuinely gravitational contribution, as also pointed out by a thought experiment in \cite{Pang}. Our Visibility expression above clarifies this by separating the contributions from the accelerated frame and gravity (tidal terms). 
    \item Since the proper time difference (denoted as $\Delta \tau$) between two paths plays a crucial role in the decoherence, one must relate $\Delta \Theta$ to $\Delta \tau$ in some manner. We write the expression for the relation between proper time and coordinate time in the appendix (\ref{Hamiltonian in FNC}) in Eq. (\ref{proper time}). The relation between $\Theta$ and $d \tau$ can be written as,
    \begin{eqnarray}\label{Theta Tau}
       d\tau = \left(1+\Theta+\frac{2}{3 m c}  R_{0 \alpha \mu \beta} x^{\alpha} x^{\beta} p^{\mu}\right)dt
    \end{eqnarray}
    \\
    The relation states that even when the proper time difference between paths is zero, it does not guarantee the absence of gravitational decoherence in general curved spacetime. 
    \item Eqs. (\ref{Theta1})-(\ref{Theta Tau}) contain only linear terms in ${a_{\mu} x^{\mu}}/{c^2}$ and reduce to Pikovski et al. \cite{Pikovski:2013qwa} results for Visibility in flat spacetime.
    If one considers $\mathcal{O}(p^2 x^2)$ terms, Eq. (\ref{Theta Tau}) becomes
    \begin{eqnarray}
        d\tau &=& \huge{\textbf{(}}1+\Theta +\frac{p^2}{ m^2 c ^2} \left(\frac{11 (a_{\mu} x^{\mu})^2}{16 c^4}\right)\nonumber \\
        &&~~~~~~~~~~~~~~~~~~~ +\frac{2}{3 m c}  R_{0 \alpha \mu \beta} x^{\alpha} x^{\beta} p^{\mu}\huge{\textbf{)}}dt 
    \end{eqnarray}
    Though the $\mathcal{O}(p^2 x^2)$ term above is by assumption small, the contribution of such a coupling term to Visibility (and hence decoherence) can not be expressed purely in terms of the proper time difference between the two superposed paths, even in the flat spacetime.

    
    \end{enumerate}
 \subsection{Curvature effect on decoherence time scale}

 It is of interest to explicitly display the decoherence time in terms of curvature. For comparison purposes, we calculate the decoherence time for the model with $N$ internal harmonic modes of the particle, considered by \cite{Pikovski:2013qwa}. The system is at rest in the superposition of two distinct positions $x_1^{\nu}$ and $x_2^{\nu}$ such that $x_2^{\nu}-x_1^{\nu}=\Delta X^{\nu}$. The internal degrees of freedom are in thermal equilibrium at local temperature $T$. The loss of Visibility is Gaussian decay for the limit $N ( k_B T \Delta \Theta / \hbar)^2 \ll 1$ with  the decoherence time 
 \begin{eqnarray}\label{Decoherence time}
     t_{\rm dec}=\sqrt{\frac{2}{N}} \frac{ \hbar c^2}{k_B T( a_{\nu} + R_{0 \mu 0 \nu} X^{\mu} )\Delta X^{\nu}}
 \end{eqnarray}
where $X^{\mu}=(x_1^{\mu}+x_2^{\mu})/2$, $k_B$ is the Boltzmann constant. The decoherence time tends to infinity, i.e., no loss in Visibility, for the case when either $\sqrt{N}k_B T$ or $ ( a_{\nu} + R_{0 \mu 0 \nu} X^{\mu} )\Delta X^{\nu}$ (or both) goes to zero. $\sqrt{N}k_B T$ is nothing but the variance in the internal energy that will go to zero if the internal states are the eigenstates of the internal Hamiltonian, i.e., $\langle H_{\rm int}^2 \rangle = \langle H_{\rm int} \rangle ^2$ and  $( a_{\nu} +R_{0 \mu 0 \nu} X^{\mu} )\Delta X^{\nu}$ is the difference in the proper time between two paths. It is obvious from the expression that the decoherence time scale (ignoring the contribution from the magnetic part of Riemann) is non-zero even when $a^\mu=0$.

We will now compare how the numerical estimate for the decoherence time scale given by Pikovski et al. in \cite{Pikovski:2013qwa} can be rederived in terms of the Riemann tensor. This turns out to be an interesting exercise. In \cite{Pikovski:2013qwa}, the idea was to consider the system on Earth, and mimic gravity by considering an accelerated frame of reference in flat spacetime, with acceleration $GM_\oplus/R_\oplus^2=9.8 \; {\rm m}/{\rm s}^2$. The decoherence time scale then evaluates to $t_{\rm dec} \sim 10^{-3}$ sec, for a system at room temperature with $N \sim 10^{23}$ and superposition size $\Delta x=10^{-6} {\rm m}$. However, it is more natural in our setup to set $a^\mu=0$ since the Earth, which will determine our Fermi frame, is, in fact, moving on a geodesic. We need to determine the Riemann tensor at the center $r=0$ of the Earth (which is our Fermi reference curve). If we ignore sources of curvature other than the Earth itself, we can evaluate this curvature from some information about the interior geometry of the Earth. Although a general analysis can be made, it will suffice for our purpose to model this as a spherically symmetric, constant density solution to Einstein equations $G_{ab}=8 \pi G T_{ab}$, given by \cite{Padmanabhan:2010zzb}:
\begin{eqnarray}
  ds^2=-f(r)dt^2+\frac{dr^2}{g(r)}+r^2 d\Omega^2   
\end{eqnarray}
where $r$ is the radial distance from the center of the constant density object with a total mass of $M$ and the radius is $R$.  The above metric satisfies the Einstein equation with $f(r)$ and $g(r)$ given by
\begin{eqnarray}
    &&f(r)=\frac{1}{4}\left[3\left(1-\frac{2GM_\oplus}{R_\oplus}\right)^{1/2}-\left(1-\frac{2 G M_\oplus r^2}{R_\oplus^3}\right)^{1/2}\right]^2\nonumber\\
    &&g(r)=\left(1-\frac{2 G M_\oplus r^2}{R_\oplus^3}\right)
\end{eqnarray}
 along with the condition $R>9G M/4$ necessary to obtain a static interior solution with finite central pressure \cite{Padmanabhan:2010zzb}. The relevant component of the Riemann tensor for this metric that appears in the decoherence time scale is $R_{\hat{0} \hat{r}\hat{0}\hat{r}}$ (frame component of Riemann tensor at the center of the Earth) and evaluates to $R_{\hat{0} \hat{r}\hat{0}\hat{r}} \to {G M_\oplus}/{R_\oplus^3}$ as $r \to 0$. In fact, this is the only non-zero component at $r=0$.
Assuming the system of interest to be close to the surface of the Earth, we also have $x_1^\mu \sim x_2^\mu \sim R_\oplus$, and hence $X^{\mu}=(x_1^{\mu}+x_2^{\mu})/2 \sim R_\oplus$. We 
therefore obtain
$$
R_{\hat{0} \hat{r}\hat{0}\hat{r}} X^{\hat{r}} \sim \frac{G M_\oplus}{R_\oplus^2}
\sim 9.8 \; {\rm m}/{\rm s}^2
$$
which precisely reproduces the estimate obtained in flat spacetime using $a \sim 9.8 \; {\rm m}/{\rm s}^2$.

\begin{figure}[H]%
    {{\includegraphics[height=6.5cm, width=0.475\textwidth]{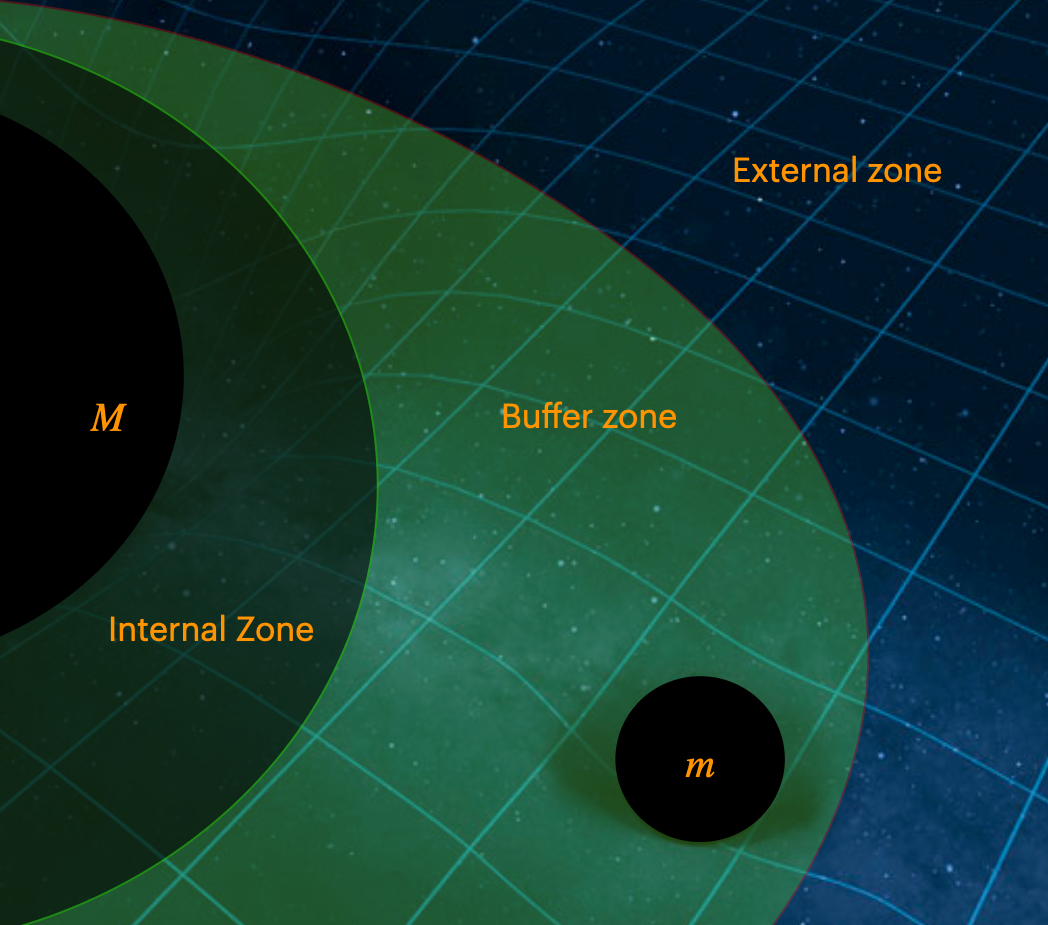} }}%
\caption{Point mass $M$ in an external environment with no matter. The internal zone (depicted in Grey) is where the gravitational field due to mass $M$ dominates the external gravitational field. The external zone (Blue colored region) is where the gravitational field due to mass $M$ is weak enough relative to external fields. The field in the buffer zone (Green colored region) interpolates between these two. Technically, if $r$ is the meaningful measure of distance from the point mass $M$ and the length scale is associated with a radius of curvature  $\mathcal{R}$ (Reimann tensor of background spacetime is inversely proportional to $\mathcal{R}^2$). The internal and external zones are defined by  $r\ll \mathcal{R}$, $r\gg M$, respectively, and the buffer zone lies between these two, where $M\ll r\ll \mathcal{R}$.}
\label{fig: BH in tidal environment}
\end{figure}

\section{Coupling of gravitational self-interaction with background curvature}
\label{Self-gravity}

To determine how the gravitational interaction itself couples to an \textit{external} curvature, we need to know how the geometry gets deformed by the system itself. This is what makes the gravitational interaction different from others, such as electromagnetic ones. While non-gravitational interactions can be accounted for by adding suitable interaction terms to the lagrangian, gravitational interaction is supposed to be incorporated in the geometry itself, which must be determined by solving Einstein equations. The question we are interested in, therefore the following: Does gravitational interaction also couple to a given, background, curvature in the same manner as, say, a point mass $m$? In this section, we will attempt to address this question by considering a simplified system comprising of two masses, $M$ and $m$, with $M \gg m$, in a given (external) curved spacetime. When $M=0, R_{abcd} \neq 0$, $m$ will couple to the background curvature $R_{abcd}$ as given in Eq. \ref{Self-H}. On the other hand, when $M \neq 0, R_{abcd}=0$, the gravitational interaction between $M$ and $m$ is incorporated through the (free) Hamiltonian of $m$ in the curvature produced by $M$; that is, Eq. \ref{Self-H} again with $g^{ab}$ given by the Schwarzschild geometry produced by $M$. To compute the coupling of the gravitational interaction between $M$ and $m$ with $R_{abcd}$, we need to consider the case When $M \neq 0, R_{abcd} \neq 0$. This is generically a hard problem in general relativity due to its non-linear structure, but solutions under various different assumptions are known.

The solution we will use derives from application of the method of matched asymptotic expansions, dating back to the work of Manasse \cite{Manasse}. Here, we will use the form of the solution derived by Poisson in \cite{Retarded} by employing retarded coordinates on the particle's world-line at $r=0$, assuming it has no acceleration in the background spacetime, and $R_{ab}=0$ (vacuum solution). The metric so derived will suffice for our purpose. To obtain its explicit form in Fermi coordinates, one can use the result of \cite{Retarded}, and apply the coordinate transformation from retarded to Fermi coordinates (carefully keeping track of higher-order terms). After a lengthy set of steps, we obtain:
\begin{widetext}
\begin{eqnarray}
   g_{t t} &=& -1-s^2 \mathcal{E^*}+M\left(\frac{2}{s}+\frac{11}{3}s\mathcal{E^*}-4 M \mathcal{E^*}\right)    
   \nonumber \\
  g_{0 \mu} &=& \frac{2}{3}s^2 \mathcal{B^*_{\mu}}+ M \Big{(}  -\frac{2}{s}\omega_{\mu}-2s \mathcal{E^*}\omega_{\mu}-\frac{2 s}{3}\mathcal{E_{\alpha \mu}}\omega^{\alpha}
  -\frac{4 s}{3}\mathcal{B^*_{\mu}}+4 M \mathcal{E^*}\omega_{\mu}\Big{)}  
   \nonumber \\
   g_{\mu \nu} &=& \delta_{\mu \nu}-\frac{1}{3}R_{\mu \alpha \nu \beta} x^{\alpha}x^{\beta}+M^2(\frac{2}{3}\mathcal{E^*_{\mu \nu}}-4 \mathcal{E^*} \omega_{\mu} \omega_{\nu})
    +M\left(\frac{2}{s} \omega_{\mu}  \omega_{\nu} +\frac{1}{3} s \mathcal{E^*}  \omega_{\mu}  \omega_{\nu}+\frac{4s}{3}\mathcal{E_{\alpha ( \mu}}\omega_{\nu )}\omega^{\alpha}+\frac{8s}{3}\mathcal{B^*_{( \mu}}\omega_{\nu)}\right)\nonumber\\
\end{eqnarray}
\end{widetext}
where
\begin{eqnarray}
        \mathcal{E}^* &=& \mathcal{E} _{\mu \nu} \omega^{\mu} \omega^{\nu}
        \\
         \mathcal{B}^*_{\alpha} &=& \epsilon_{\alpha \beta \gamma} \mathcal{B}^{\beta}_{\mu}\omega^{\mu} \omega^{\gamma} 
         \\
          \mathcal{E}^*_{\mu \nu}&=&2 \mathcal{E}_{\mu \nu}-4 \omega_{(\mu} \mathcal{E}_{\nu ) \alpha} \omega^{\alpha} +(\delta_{\mu \nu}+\omega_{\mu}\omega_{\nu})\mathcal{E}^*
\end{eqnarray}
$\mathcal{E} _{\mu \nu}$ and $\mathcal{B} _{\mu \nu}$ are symmetric traceless tensors with $\mathcal{E} _{\mu \nu}=R_{0 \mu 0 \nu}$ and $\mathcal{B}^{\delta}_{\alpha}$, defined such that $R_{\alpha 0 \beta \gamma}=\epsilon _{\beta \gamma \delta}\mathcal{B}^{\delta}_{\alpha}$. $\mathcal{B}^*_{\alpha}$ and $\mathcal{E}^*_{\mu \nu}$ are the combination of the electric and magnetic parts of the Riemann tensor such that these are orthogonal to $\omega^{\mu}$.  The metric tensor depends on Fermi time $t$ through $\mathcal{E} _{\mu \nu}(t)$ and $\mathcal{B} _{\mu \nu}(t)$ and angular dependence is encoded in $\omega(\theta^A)$.

It is straightforward to check that the above metric reduces to the one in \cite{Manasse} when the external environment itself is Schwarzschild geometry of a spherically symmetric mass distribution. The general form above also makes it clear that the curvature in the neighborhood of the composite system with mass $M$ arises due to two factors: The Schwarzschild field of $M$ itself, and the curvature produced by the external environment. Einstein equations are satisfied throughout the motion.

The term that will be of interest to us will arise from the coupled terms such as $(M \mathcal{E^*})$ above. The Hamiltonian for a free particle of mass $m$ (assumed to be at rest, for simplicity) in the buffer zone (see Fig. \ref{fig: BH in tidal environment}) using the relation (\ref{-p0}), becomes
\begin{align}\label{Self-H}
    H=m c^2\left(1+\frac{1}{2} s^2 \mathcal{E}^*\right)+H_{\rm int} \left(1+\frac{4}{3}s^2 \mathcal{E}^*+\frac{4}{3}s^2 B^{* \mu} \omega_{\mu}\right)\nonumber\\
\end{align}
where $H_{\rm int}=-G M m/s$ is the gravitational interaction between two bodies with masses $M$ and $m$. While writing the above expression, we have used the time-time component of the inverse deformed Fermi metric correct upto the order $M$,
\begin{eqnarray}
    g^{t t}=-1+s^2 \mathcal{E}^* -M\left(\frac{2}{s}+\frac{1}{3}s \mathcal{E}^*+\frac{8}{3}s B^{* \mu} \omega_{\mu}\right)
\end{eqnarray}
Note that the above Hamiltonian can be rewritten as 
\begin{align}
H = 
\left( m c^2 + H_{\rm int} \right) \left(1+\frac{1}{2} s^2 \mathcal{E}^*\right)
+ H_{\rm int} \left(\frac{5}{6}s^2 \mathcal{E}^*+\frac{4}{3}s^2 B^{* \mu} \omega_{\mu} \right) 
\end{align}
which is not equivalent to replacing $m \to m + H_{\rm int}/c^2$ in the single particle Hamiltonian, due to the second term.

It has been mentioned in the literature that such an asymmetric coupling, for non-gravitational interactions, is a coordinate artifact and disappears when covariance is maintained \cite{PhysRevA.100.052116, Zych}. However, our analysis has been covariant all along, and hence the above effect is real. This can be stated more rigorously, by noticing that the $1/r$ terms that appear in the interaction Hamiltonian are already covariantly defined in terms of retarded coordinate $r$: 
$$r=\left[ u^a \nabla_a \sigma \right]_{\rm ret}$$

We can rewrite the above Hamiltonian in retarded coordinates by using the coordinate transformation given in \cite{Poisson:2011nh}, to obtain 
\begin{align}
    &H = m c^2\left(1+\frac{1}{2} r^2 \mathcal{E}^*\right)+H_{\rm int} \left(1+\frac{3}{2}r^2 \mathcal{E}^* +\frac{4}{3}r^2 B^{* \mu}\Omega_{\mu}\right)
    \nonumber\\
  &= \left( m c^2 + H_{\rm int}\right) \left(1+\frac{1}{2} r^2 \mathcal{E}^*\right)+H_{\rm int} \left(r^2 \mathcal{E}^* +\frac{4}{3}r^2 B^{* \mu}\Omega_{\mu}\right)
\end{align}
where now $H_{\rm int}=-GMm/r$ is the interaction Hamiltonian defined using the correct, retarded, coordinate $r$, and the curvature tensor(s) $B^{* \mu}(u)$, $\mathcal{E}^*(u)$ are now functions of the retarded time $u$. Once again, the above Hamiltonian is not equivalent to replacing $m \to m + H_{\rm int}/c^2$ in the single particle Hamiltonian, due to the second term.

\section{Discussions}
\label{discussions}

Amongst all interactions, the gravitational interaction is the most ubiquitous because it is encoded in the curvature of spacetime itself. Being so, it couples to everything and operates \textit{un-shielded} with an infinite range. This is what makes the study of quantum systems in a gravitational field subtle as well as interesting, yielding results that can be as interesting as those that might come from quantum gravity in which gravity and spacetime themselves are treated quantum mechanically. Ever since the seminal work of Pikovski et al., the universal role of gravity in quantum effects such as entanglement and decoherence has been studied extensively. However, while research along these lines continues to be active, there have been arguments raised and issues of concept highlighted in various discussions \cite{Pang, Bonder:2015hja, Bonder:2015tma, Diosi:2015vra}. In this work, our aim was to settle some of the key amongst these issues:
\begin{enumerate}
    \item Does the decoherence effect conflict with the equivalence principle?
    \item Will there be decoherence in freely falling frames?
    \item Does the gravitational self-interaction $H_{\rm int}$ couple to the external curvature through the simple replacement $m \to m + H_{\rm int}/c^2$?
\end{enumerate}
Our analysis settles the answer to these as (1.) No, (2.) Yes, and (3.) No.

Most previous debates and discussions surrounding the above questions have been based on arguments that are qualitative, and, therefore, can not provide any insights into the role of curvature in answering any of them. Our results do precisely that, and in the process, we obtain additional insights that could not have been obtained from qualitative arguments alone.

Let us start with the role of the equivalence principle. In fact, different versions of this principle appear in (1) and (3) above. In (1), the essential question is whether results in an accelerated frame can truly mimic gravity, and/or capture curvature effects. In our opinion, analysis in an accelerated frame of reference in flat spacetime can, at best, act as an indicator of the \textit{existence} of certain phenomenon in gravitational fields, and nothing more. Precise quantitative results in a gravitational field can generically not be estimated/obtained by working in an accelerated frame in flat spacetime. Although there are examples that might be considered as exceptions, such as the mapping between the Unruh effect and Hawking radiation, even in such cases, curvature might play an important role otherwise invisible to perturbative analysis, as has been recently pointed out in \cite{K:2021gns}. In this work, it was shown that the response of the conventional Unruh-deWitt detector accelerating in a direction $\bf n$, in a curved spacetime, will generically have a thermal part that corresponds to a temperature of $T = (\hbar/2\pi) \sqrt{a^2 - \mathscr{E}_{n}}$ (where $\mathscr{E}_{n}=R_{0n0n}$), \textit{with no restriction on the relative strength of $a$ and $\mathscr{E}_n$}. This is surprising since we are dealing with a point detector (hence no tidal forces), and yet the contribution of at least one component of curvature can not be fixed by appealing to the equivalence principle. A dramatic illustration of this result obtains in (anti-)de Sitter spacetimes, where the result is exact and reproduces the known Unruh temperature in these spacetimes.

The above form of the equivalence principle is, therefore, operative in a very restricted sense. Such pitfalls related to the above form of the equivalence principle have been noted in the past, most prominently by Synge \cite{Synge-Book}. Nevertheless, we were able to show that quantitative estimates obtained using the Riemann tensor do agree with the ones using accelerated frames in PZCB, at least for experiments done near a spherically symmetric source. This is a curious result that has not been noted in the literature so far, to the best of our knowledge, and deserves further thought. 

The answer to point (2) above follows immediately from the fact that gravitational or curvature-induced decoherence appears via the term $R_{0 \mu 0 \nu} X^{\mu}$, which is generically non-zero in any frame, including an inertial one. In fact, it can be interpreted as the acceleration associated with {\it deviation} between two curves; for $a^i=0$, the relevant curve is the geodesic $\gamma$, and the curve defined by $X^{\mu}=(x_1^\mu+x_2^\mu)/2$. We must hasten to add, however, that the curve so defined would, in general, not be a geodesic, and hence an interpretation purely in terms of deviation acceleration would break down. Given its elegance, though, it would be interesting to pursue this line of thought further. Furthermore, this discussion also addresses the issue of whether the decoherence derived by Pikovski et al. \cite{Pikovski:2013qwa} is actually kinematic and can be nullified in certain frames of reference \cite{Pang}. Interestingly, there will be no loss of coherence even for a uniformly accelerated observer, such as the one considered by Pikovski et al., when $a_{\mu}$ is such that it exactly cancels the term $R_{0 \mu 0 \nu} X^{\mu}$. To summarize, our analysis clearly identifies that decoherence will generically exist in curved spacetime, but the decoherence time scale $t_{\rm dec}$ will quantitatively depend on the frame of reference because (i) the relevant curvature component $R_{0 \mu 0 \nu}=R_{abcd} u^a {\sf e}^b_\mu u^c {\sf e}^d_\nu$ depends on the frame, and (ii) the vector $X^{\mu}$ depends on the choice of frame. Note that generically the component in (i) will be non-zero in curved spacetime, while an interesting case when the curvature term in 
$t_{\rm dec}$ disappears is in a frame in which $X^\mu=0$, that is $x_1^\mu=-x_2^\mu$. Nevertheless, as long as the magnetic part of Riemann is zero, and/or the momentum coupling is ignored, the Visibility will depend on the proper time difference of the superposed paths, and hence frame independent.

On the other hand, the version of the equivalence principle that appears in (3) is the strong one, which in this case would mean that the coupling of gravitational interaction to external background curvature can be determined in the same universal manner as coupling of non-gravitational interactions is determined. Our analysis shows that this is not true, and hence one can not use the replacement $m \to m + H_{\rm int}/c^2$ when $H_{\rm int}$ is the self-gravity of the system. Once again, such a result could not have been guessed from qualitative arguments alone and seems to be in conflict with the results in \cite{Zych}. Unlike non-gravitational interactions, gravitational interactions are encoded in the geometry of spacetime, and hence their coupling to the external field is constrained by gravitational field equations. Note that, unlike in \cite{PhysRevA.100.052116, Zych}, our result is derived using a proper solution of general relativity rather than in the Newtonian limit and employs variables that are covariantly defined. Even the ``radial" coordinate $r$ is defined covariantly (and is the affine parameter along retarded null geodesics). We would like to add that, strictly speaking, the strong equivalence principle holds when tidal effects can be ignored, and since the violations we have arise from tidal terms, they are not in conflict with the strong equivalence principle. Nevertheless, it does highlight the non-trivial difference between gravitational and non-gravitational interactions.
\\
\\
\noindent \textbf{Acknowledgement:} The authors would like to thank Prof. Alessandro Pesci for a useful correspondence. DK acknowledges the support from the Institute of Eminence scheme of IIT Madras funded by the Ministry of Education of India.

\appendix
\section{Free particle Hamiltonian in curved spacetime}
 \label{Hamiltonian in FNC}
We write the metric in Fermi normal coordinate \cite{Manasse:1963zz, Poisson:2011nh} near an accelerated curve $\gamma$ with $a_{\mu}$ are the components of the acceleration vector and described by $x^{\alpha}=0$.
\begin{equation} \label{Fermi00}
    g_{0 0}=-\left( 1+\frac{2}{c^2}a_{\mu}x^{\mu}+(\frac{a_{\mu}x^{\mu}}{c^4})^2+R_{0 \mu 0 \nu}x^{\mu}x^{\nu} \right)+\mathcal{O}(s^3),
 \end{equation}
\begin{equation}
\label{Fermi0mu}
     g_{0 \mu}=-\frac{2}{3}R_{0 \alpha \mu  \beta}x^{\alpha} x^{\beta}+\mathcal{O}(s^3),
\end{equation}   
   
\begin{equation}
    g_{\mu \nu}=\delta_{\mu \nu}-\frac{1}{3}R_{\mu \alpha \nu \beta}x^{\alpha}x^{\beta}+\mathcal{O}(s^3).
\end{equation}
where the components of the Riemann tensor are evaluated on the curve $\gamma$. Here we write the point particle Hamiltonian($H_{\rm Free}$) in the Fermi normal coordinates, which is uplifted to the Hamiltonian for the composite system using the center of mass degree of freedom. We adopted the definition of the Hamiltonian $H$ for a point particle in FNC as $H=-p_0$ (\cite{Kothawala:2011fm}), which can  further be written explicitly in terms of metric as,
\begin{equation}
   H= \frac{g^{0\mu}p_{\mu}c}{g^{00}} + \sqrt{\frac{g^{\mu\nu}p_{\mu}p_{\nu}c^2+m^2 c^4}{-g^{00}} + \bigg(\frac{g^{0\mu}p_{\mu}c}{g^{00}}\bigg)^2}
   \label{-p0}
\end{equation}
\begin{widetext}
    \begin{eqnarray}
        H_{\rm Free}-m c^2&=&+\frac{2}{3}R_{0 \alpha \ \beta }^{\ \ \mu } x^{\alpha} x^{\beta} p_{\mu}c+\frac{\vec p ^2}{2 m}\left(1+\frac{ a_{\mu} x^{\mu}}{ c^2}+\frac{15(a_{\mu} x^{\mu})^2}{8 c^4} +\frac{R_{0 \mu 0 \nu}x^{\mu}x^{\nu}}{2}\right)+\frac{m c^2}{2}\left(\frac{2 a_{\mu} x^{\mu}}{ c^2}+R_{0 \mu 0 \nu}x^{\mu}x^{\nu}\right)\nonumber \\
        &&+\frac{1}{6}R^{\mu \ \nu}_{\ \alpha \ \beta} x^{\alpha}x^{\beta}\frac{p_{\mu}p_{\nu}}{m} 
    \end{eqnarray}
Hamiltonian $H$ for a system whose internal dynamics are governed by the Hamiltonian $H_{\rm int}$,

\begin{eqnarray} \label{H}
    H &=&H_{\rm Free} +H_{\rm int}\huge{\textbf{(}}1-\frac{\vec p  ^2}{2 m^2 c^2}\left(1+\frac{ a_{\mu} x^{\mu}}{ c^2}+\frac{15(a_{\mu} x^{\mu})^2}{8 c^4}  +\frac{R_{0 \mu 0 \nu}x^{\mu}x^{\nu}}{2}\right)
    +\frac{1}{2}\left(\frac{2 a_{\mu} x^{\mu}}{ c^2}+R_{0 \mu 0 \nu}x^{\mu}x^{\nu}\right)\nonumber \\
    &&~~~~~~~~~~~-\frac{1}{6}R^{\mu \ \nu}_{\ \alpha \ \beta} x^{\alpha}x^{\beta}\frac{p_{\mu}p_{\nu}}{m^2 c^2} \huge{\textbf{)}}
\end{eqnarray}
An infinitely small change in  proper time can be written in terms of Fermi time as,
    \begin{eqnarray}\label{proper time}
        d \tau &=& d t \huge{\textbf{(}}1+\frac{ a_{\mu} x^{\mu}}{ c^2}+\frac{R_{0 \mu 0 \nu}x^{\mu}x^{\nu}}{2} -\frac{\vec p  ^2}{2 m^2 c^2}\left(1-\frac{ a_{\mu} x^{\mu}}{ c^2}+\frac{(a_{\mu} x^{\mu})^2}{2 c^4} -\frac{R_{0 \mu 0 \nu}x^{\mu}x^{\nu}}{2}\right)+\frac{2}{3} R_{0 \alpha \mu \beta}\frac{x^{\alpha} x^{\beta} p^{\mu}}{m c}\nonumber \\
         &&~~~~~~~~~~~+\frac{1}{6}R^{\mu \ \nu}_{\ \alpha \ \beta} x^{\alpha}x^{\beta}\frac{p_{\mu}p_{\nu}}{m^2 c^2} \huge{\textbf{)}}
    \end{eqnarray}
\end{widetext}

\bibliography{apssamp}
\end{document}